\documentclass{nature}

\usepackage{float}
\usepackage{caption}
\usepackage[displaymath, mathlines]{lineno}
\usepackage{graphicx,amssymb}
\usepackage{subfigure,appendix,enumerate}
\usepackage{url,hyperref,xcolor}
\usepackage{multirow}
\usepackage{lscape}
\usepackage{epsfig}
\usepackage[latin9]{inputenc}
\usepackage{rotating}
\usepackage{journal-names}
\usepackage{color}
\usepackage{epstopdf}

\def\simmore{\mathbin{\lower 3pt\hbox
     {$\rlap{\raise 5pt\hbox{$\char'076$}}\mathchar"7218$}}} 

\newcommand{\comment}[1]{}

\title{The X-ray corona turns into the relativistic jet in the micro quasar GRS 1915+105} % 

\author{Mariano M\'endez$^{*,1}$, Konstantinos Karpouzas$^{1,2}$, Federico Garc\'{\i}a$^{1,3}$, Liang Zhang$^{2}$, Yuexin Zhang$^{1}$, Tomaso M. Belloni$^{4}$ \& Diego Altamirano$^{2}$}
\usepackage[backend=biber,style=nature,articletitle=true,intitle=true,natbib=true, citecounter=true]{biblatex}

\usepackage{totcount}
\newtotcounter{citenum} 

\makeatletter
\AtEveryCitekey{%
  \ifcsundef{blx@entry@refsegment@\the\c@refsection @\thefield{entrykey}}
    {\csnumgdef{blx@entry@refsegment@\the\c@refsection @\thefield{entrykey}}{\the\c@refsegment}}
    {}}
\defbibcheck{onlynew}{%
  \ifnumless{0\csuse{blx@entry@refsegment@\the\c@refsection @\thefield{entrykey}}}{\the\c@refsegment}
    {\skipentry}
    {}}
\makeatother

\makeatletter
\AtDataInput{%
  \ifltxcounter{datacount@\the\c@refsection}
    {}
    {\newcounter{datacount@\the\c@refsection}}%
  \stepcounter{datacount@\the\c@refsection}}

\newcommand*{\refsectionbibcount}{\arabic{datacount@\the\c@refsection}}
\makeatother

\usepackage{xstring}

\newwrite\tempfile

\begin{document}

%TC:ignore

\maketitle

\begin{affiliations}
\item Kapteyn Astronomical Institute, University of Groningen, Postbus 800, 9700 AV Groningen, The Netherlands 
\item School of Physics and Astronomy, University of Southampton, Southampton, SO17 1BJ, UK
\item Instituto Argentino de Radioastronom\'{\i}a (CCT La Plata, CONICET; CICPBA; UNLP), C.C.5, (1894) Villa Elisa, Buenos Aires, Argentina
\item INAF - Osservatorio Astronomico di Brera, Via E. Bianchi 46, I-23807 Merate, Italy
\end{affiliations}

%TC:endignore

\newrefsegment

\begin{linenumbers}

\begin{abstract}

%!TEX root = GRS1915.tex 
\linenumbers
GRS~1915+105\autocite{Fender-2004a} was the first stellar-mass black-hole in our Galaxy to display a superluminal radio jet\autocite{Mirabel-1994}, similar to those observed in active galactic nuclei with a supermassive black hole at the centre\autocite{Pearson-1981}.
It has been proposed that the radio emission in GRS~1915+105 is fed by instabilities 
in the accretion disc\autocite{Belloni-1997b} by which 
the inner parts of the accretion flow
is ejected in the jet\autocite{Mirabel-1998, Eikenberry-1998, Fender-2004b}.
Here we show that there is a significant correlation between: (i) the radio flux, coming from the jet, and the flux of the iron emission line, coming from the disc and, (ii) the temperature of the corona that produces the high-energy part of the X-ray spectrum via inverse Compton scattering
and the amplitude of a high-frequency variability component coming from the innermost part of the accretion flow. At the same time, the radio flux and the flux of the iron line are strongly anti-correlated with the temperature of the X-ray corona and the amplitude of the high-frequency variability component. These correlations persist over $\sim$$10$ years, despite the highly variable X-ray and radio properties of the source in that period\autocite{Belloni-2000, Pooley-1997}. 
Our findings provide, for the first time, incontrovertible evidence that the energy that powers this black-hole system can be directed either to the X-ray corona or the jet. When this energy is used to power the corona, raising its temperature, there is less energy left to fuel the jet and the radio flux drops, and vice versa. 
These facts, plus the modelling of the variability in this source show conclusively that in GRS~1915+105 the X-ray corona morphs into the jet.\\

\end{abstract}

The X-ray spectrum of black-hole binaries can be decomposed into three main radiation components: (i) A thermal component that dominates the emission at low (soft) energies\autocite{Shakura-1973}, below $\sim$$5$ keV, due to an accretion disc through which mass flows from the binary companion to the black hole. (ii) A power-law like component that dominates the spectrum at energies above $\sim$$5$$-$$10$ keV, caused by inverse-Compton scattering of the disc photons in a corona of highly-energetic electrons\autocite{Sunyaev-1980}. If the energy distribution of the electrons is Maxwellian, this component features a high-energy cut off at an energy that depends upon the electron temperature of the corona, $kT_{\rm e}$. (iii) A broad emission line at $\sim$$6.5$$-$$7$ keV due to iron, produced when photons from the corona reflect off the disc, with the line profile being set by special- and general-relativistic effects\autocite{Fabian-1989, Fabian-2009}.
The X-ray emission of accreting black-hole binaries exhibits high-amplitude variability from tens of milliseconds to decades\autocite{Greiner-1996, Belloni-2000}. Depending on the relative importance of these spectral components and the strength of the variability, accreting black holes display different states\autocite{Mendez-1997}: In the hard states the X-ray spectrum ($\sim$$1$$-$$20$ keV range) is dominated by emission from the corona, and the Fourier power spectrum shows variability of up to 50\% of the average luminosity over a broad range of time scales plus narrow quasi-periodic oscillations, QPOs. In the soft states the emission is dominated by the accretion disc, and the variability drops to less than 5\% of the average luminosity.  

Black-hole binaries in the hard states emit in radio with a spectrum that is consistent with self-absorbed synchrotron radiation from an optically thick and compact jet\autocite{Fender-1997}.  
During the transition from the hard to the soft states, some black-hole binaries show radio emission from individual, spatially-resolved, plasma clouds that are ejected in a jet at speeds close to the speed of light. The radio spectrum of these discrete ejections is consistent with synchrotron emission from optically-thin material\autocite{Mirabel-1994}.

GRS 1915+105 harbours a $12^{+2.0}_{-1.8}$ solar-mass black hole\autocite{Reid-2014} and is very variable both in X-rays and radio wavelengths\autocite{Belloni-2000, Pooley-1997}. In X-rays, the emission switches from times in which a bright accretion disc with a temperature $kT_{\rm bb}$$\approx$$2$ keV and small inner-radius dominates the spectrum, to times in which the corona dominates the spectrum, the disc is relatively cool, $kT_{\rm bb}$$\approx$$0.5$$-$$1$ keV and is inferred to have a large inner radius\autocite{Belloni-1997b}. In several observations, mostly those with a cool disc, in addition to a band-limited noise component, a narrow and strong QPO, called type-C QPO\autocite{Casella-2005}, appears in the Fourier power spectrum at frequencies between $\sim$$0.4$ Hz and $6.5$ Hz \autocite{Zhang-2020}. The frequency of the QPO increases as the temperature of the disc increases and the spectrum of the source softens. Besides this QPO (and harmonics and sub-harmonics of the fundamental frequency), the power spectrum of GRS 1915+105 sometimes shows a broad variability component at $\sim$$60$$-$$80$ Hz\autocite{Trudolyubov-2001} that we will call  the high-frequency bump. The high-frequency bump appears when the spectrum of the source is dominated by  the corona, but observations in which the corona dominates the emission do not always show this high-frequency bump. At the same time, observations of GRS 1915+105 in which the spectrum  is dominated by  the corona are sometimes, but not always, accompanied by high radio fluxes\autocite{Muno-1999, Fender-1999}. (We show the power-density spectra of two observations of GRS 1915+105 with the QPOs and the bump indicated in the Extended Data Figure~\ref{fig:power-spectra}.)

We studied a large dataset of 1800+ X-ray observations of GRS 1915+105 obtained with the Rossi X-ray Timing Explorer (RXTE) between 1996 and 2012, combined with almost daily observations of the source at 15 GHz with the Ryle telescope. Our final sample consists of $410$ observations for which we have simultaneous data both in X-rays and radio. For each of these observations we have a measurement of (i) the radio flux density at 15 GHz, (ii) the X-ray hardness ratio calculated as the ratio of the intensity in the $13$$-$$60$ keV to the $2$$-$$7$ keV band, (iii) the frequency of the fundamental component of the type-C QPO, (iv) the phase lag at the QPO frequency for photons in the $5.7$$-$$15$ keV band with respect to those in the $2$$-$$5.7$ keV band, (v) the $2$$-$$60$ keV fractional rms amplitude of the high-frequency bump, and (vi) the best-fitting parameters to the X-ray energy spectra of the source. (See the section Methods for details of the analysis and an explanation of some of these quantities). These vastly different types of data, consisting of X-ray and radio fluxes and spectral and timing properties of a single source, come from wavelengths that are more than eight orders of magnitude apart and sample time scales that are more than eleven orders of magnitude different, from ten milliseconds to a decade.

In Figure~\ref{fig:fig1} we plot the X-ray hardness ratio as a function of the frequency of the QPO for these $410$ observations of GRS~1915+105. The $x$ and $y$ axes of the four panels in this Figure are the same, while the colours of the points in each panel represent, respectively, the simultaneous 15-GHz radio-flux measurements from the jet (Fig.~\ref{fig:fig1a}), the electron temperature of the X-ray corona (Fig.~\ref{fig:fig1b}), the flux of the iron emission line in the X-ray spectrum (Fig.~\ref{fig:fig1c}), and the fractional rms amplitude of the high-frequency bump (Fig.~\ref{fig:fig1d}). In all four panels blue (red) indicates high (low) values of the quantity represented by the colours.

The QPO frequency generally increases as the hardness ratio decreases and the source spectrum softens, consistent with a decreasing inner radius of the accretion disc that leads to an increase of both the disc flux\autocite{Shakura-1973} and QPO frequency\autocite{Stella-1998, Ingram-2009}. The relation, however, is significantly broader than the spread expected from the errors in each quantity. (The errors are smaller than the size of the points.)

The colours of the points in the four panels of this Figure show that the radio flux, the flux of the iron line, the temperature of the corona and the fractional rms amplitude of the high-frequency bump depend upon both QPO frequency and hardness ratio. This double dependence is the reason for the striking match between the two panels on the left and the two on the right, and the almost perfect mirror match between the two top and the two bottom panels. In other words, the breadth of the hardness-ratio/QPO-frequency relation is consistently set by either of the four quantities, such that if we plotted the data in a diagram with the QPO frequency along the $x$ axis, the hardness ratio along the $y$ axis and any of these four quantities along the $z$ axis, the points would lie on a 2-dimensional surface in 3D. 

The two panels of Figure~\ref{fig:gauss-flux} show the flux of the iron line as a function of the X-ray flux in the $2$$-$$25$-keV band for all observations of GRS~1915+105. In the left panel the colour of the points represents the temperature of the corona, while in the right panel the colours represent the phase lags between the $2$$-$$5.7$ keV and $5.7$$-$$15$ keV bands at the frequency of the QPO. While the flux of the iron line increases as the total flux of the source increases, when the temperature of the corona is low, the radio flux is high and the lags at the QPO frequency are positive the correlation\autocite{Neilsen-2009} is steeper than when the temperature is high, the radio flux is low and the lags are negative. To confirm that our measurements of the line flux are not biased because of the relatively low spectral resolution of the RXTE/PCA instrument, we also include in this Figure independent measurements of the flux of the line with Chandra and Suzaku taken from the literature\autocite{Neilsen-2009, Mizumoto-2016} (black plus and cross symbols).

These results provide a unique clue to understand the nature of the corona and the jet in this object, and what powers these components. Inverse Compton scattering cools down the corona by transferring energy from the electrons to the soft disc photons. The temperature of the corona in black-hole binaries, however, increases during periods in which the photon flux of the disc drops and Compton cooling is less effective. This means that a source of power balances the inverse Compton cooling and sets the temperature of the corona. Our findings show that in GRS 1915+105 the energy provided by this mechanism is split to either power the jet or heat the corona. 

The thermal energy stored in a spherical corona of optical depth $\tau$ and size $L$ around a black hole of mass $M$ is $E_{th}$$\approx$$3.4$$\times$$\displaystyle 10^{26} \tau \left(\frac{kT_e}{1\;\mathrm{keV}}\right) \left(\frac{M}{M_\odot}\right)^2 \left(\frac{L}{R_g}\right)^2$ erg, where $R_g=GM/c^2$ is the gravitational radius of the black hole and $G$ and $c$ are, respectively, the gravitational constant and the speed of light\autocite{Merloni-2001}. 

As discussed in the section Extended Results and Discussion, the change of sign of the lags is explained if a fraction of the disc photons that are Comptonised in the corona impinges back onto the disc before reaching the observer\autocite{Karpouzas-2020}. When this feedback fraction is low, time delays due to Comptonisation dominate and the lags are positive; when this fraction is high, reprocessed disc photons reach the observer later than those from the corona and the lags become negative. Taking the corona sizes from fits with this model\autocite{Karpouzas-new}, $L$$\approx$$10$$-$$1200 R_g$, plus $\tau$$\approx$$1$$-$$6$ and $kT_e$$\approx$$5$$-$$40$ keV from the spectral fits, we find that $E_{th}$$\approx$$10^{31}$$-$$10^{35}$ erg. If this energy is released over the time scale of the high-frequency bump\footnote{Note that this is the shortest variability time scale in the data; using any other time scale longer than this one to estimate the thermal luminosity would make the discrepancy even bigger.}, the thermal luminosity is $2$ to $5$ orders of magnitude lower than the observed luminosity of the corona in GRS 1915+105. The alternative is that the corona is powered by magnetic energy\autocite{Merloni-2001, Malzac-2004}, e.g., shear energy due to differential rotation of the magnetic-field lines that thread the accretion disc. This magnetic energy would also be responsible for the synchrotron radio emission and the jet ejection mechanism in black-hole binaries.

We propose that in GRS~1915+105 the corona turns into the jet and that both are, at different times, the same physical component. Based on the results shown here and fits with the model of the lags that we present in the section Extended Results and Discussion, the process proceeds as follows: 
(1) When the QPO frequency is $\sim$$6$ Hz the corona is relatively  large and enshrouds the inner parts of the accretion disc (Fig.~\ref{fig:fig3c}a); as seen from the corona, the disc covers half of the sky and hence there is a high probability that photons from the corona return to the disc leading to a high feedback fraction and negative QPO lags. As the QPO frequency decreases from $\sim$$6$ Hz to $\sim$$2$ Hz, the inner edge of the disc moves outwards and the magnitude of the lags and the size of the corona decrease. In this phase the magnetic field that threads the disc and powers the corona is relatively disorganised, the magnetic energy is dissipated stochastically leading to a high corona temperature. 
(2) At a QPO frequency\footnote{The transition occurs when the lags fo the QPO change from negative to positive; although the exact value of the QPO frequency\autocite{Zhang-2020} at which this transition occurs is between 1.8 and 2 Hz, here we will always write 2 Hz for simplicity.} of $\sim$$2$ Hz the size of the corona becomes equal to the truncation radius of the accretion disc, the feedback fraction decreases abruptly and the QPO lags become zero. From this point on the corona moves inside the accretion disc (Fig.~\ref{fig:fig3c}b), the magnetic field starts to become coherent on longer spatial scales, the stochastic energy dissipation decreases and the temperature of the corona drops. At the same time the magnetic-field lines start to channel material from the corona into the direction perpendicular to the accretion disc, and low-level radio emission from the jet starts to appear.   
(3) As the QPO frequency continues decreasing below $\sim$$2$ Hz the inner disc radius moves further out and the size of the corona increases again but, given that the lags are positive, this time the feedback fraction remains low. From this it follows that the corona does not cover the inner parts of the disc, which can only be the case if the geometry of the corona changes such that it becomes more prominent in the direction perpendicular to the disc (Fig.~\ref{fig:fig3c}c); as seen from the corona, the disc now covers a much smaller area of the sky. As the magnetic-field lines become more spatially coherent\autocite{Meier-2005} there is less stochastic energy dissipation, the corona temperature drops further and the material from the corona that was channeled off the originally extended corona is expelled away from the accretion-disc plane and becomes the radio jet (Fig.~\ref{fig:fig3c}d).

Our multi-wavelength correlations match the proposal that, during the initial parts of an outburst, the X-ray corona of the black-hole binary MAXI J1820+070 contracts\autocite{Kara-2019} and then re-expands\autocite{Wang-2021}. Here we show conclusively that, as previously speculated\autocite{Vadawale-2003, Rodriguez-2003, Fender-2004b}, in the case of GRS 1915+105 the contracting corona becomes the radio jet and that, at least part of the time, the corona and the jet are actually one and the same physical component\autocite{Levinson-1996, Giannios-2004, Markoff-2005}.

The appearance of radio flares when the QPO frequency is below 2 Hz reinforces our interpretation. Figure~\ref{fig:fig2} shows that in the periods that the frequency of the QPO moves more or less stochastically between $\sim$$2$ Hz and $\sim$$8$ Hz (red circles) the radio flux (light blue curve) is low. Occasionally, the QPO frequency evolves in a more systematic way: it starts to decrease, crosses the value of 2 Hz and moves up again; at the same time the radio flux increases sharply and a radio flare lasting a few tens of days is observed. Figure~\ref{fig:fig2} shows this behaviour over a period of about 500 days in which the source shows two radio flares. Extended Data Figure~\ref{fig:radio-qpo} shows that the same behaviour repeats consistently over a period of 10 years and about a dozen radio flares. 

The relation between the iron-line flux and the total flux is consistent with the above scenario. The flux of the iron line depends more strongly upon X-ray flux when the temperature of the corona is low, that is when the corona has turned into the jet\autocite{Neilsen-2009}, than when the corona is more extended and covers the inner parts of the accretion disc (Fig.~\ref{fig:gauss-flux}) . This points to a lamp-post geometry\autocite{Miniutti-2004} with the corona, which is now the jet, illuminating the disc anisotropically\autocite{Kylafis-2020, Reig-2021}. Because the jet does not cover the accretion disc, the flux of the iron line is not (or mildly) attenuated by the corona\autocite{Matt-1997, Petrucci-2001b}.

The behaviour we observe in GRS~1915+105 could explain the deviations from a single track in the radio/X-ray correlation of other accreting galactic black
holes\autocite{Hannikainen-1998, Corbel-2003, Gallo-2003, Coriat-2011, Gallo-2012, Gallo-2018}
If, as in the case of GRS 1915+105, the energy powering these systems is used to either accelerate the jet or heat the corona, different sources, or the same source at different times, will show 
lower radio fluxes at a given X-ray flux (or, equivalently, higher X-ray fluxes at a given radio flux), depending on how much energy is directed towards, respectively, launching the radio jet or heating the X-ray corona.

\end{linenumbers}

\printbibliography[segment=\therefsegment,check=onlynew]

%TC:ignore

\newpage
\begin{addendum}
\itemsep0em 
 \item This work is part of the research programme Athena with project number 184.034.002, which is (partly) financed by the Dutch Research Council (NWO). Y.Z. acknowledges the support from China Scholarship Council Scholarship (201906100030). TMB acknowledges financial contribution from the agreement ASI-INAF n.2017-14-H.0, PRIN-INAF 2019 N.15, and thanks the Team Meeting at the International Space Science Institute (Bern) for fruitful discussions. D.A acknowledges support from the Royal Society. O.B. and D.A. acknowledge support of a Diamond Jubilee International Visiting Fellowship by the University of Southampton. We thank G. Pooley for making the radio data available. This research has made use of data and/or software provided by the High Energy Astrophysics Science Archive Research Center (HEASARC), which is a service of the Astrophysics Science Division at NASA/GSFC. This research made use of NASA's Astrophysics Data System.

\item[Author Contributions]
All authors contributed to interpretation of results and edited the manuscript. M.M. led the interpretation, obtained spectral parameters and wrote the manuscript. K.K. wrote the model that triggered this research, produced initial radio/timing plots, fitted rms and lag spectra and co-led the interpretation. F.G. produced initial 3D radio/timing/spectral plots, fitted rms and lag spectra of the QPO and co-led the interpretation. M.M., K.K. and F.G. measured extra QPO frequencies. L.Z. obtained parameters of the QPO. Y.Z. obtained parameters of the high-frequency bump.  T.M.B. had the idea to study the high-frequency bump in connection with the radio flux. D.A. and O.B. discussed the results and contributed to the interpretation.

\item[Competing Interests] The authors declare that they have no competing financial interests.\\
Supplementary Information is available for this paper.
\item[Correspondence] Correspondence and requests for materials
should be addressed to MM\\(email: mariano@astro.rug.nl).
Reprints and permissions information is available at www.nature.com/reprints

\end{addendum}

\newrefsegment

\newpage
%!TEX root = GRS1915.tex 
\renewcommand{\figurename}{} 
\begin{minipage}{\linewidth}
\renewcommand{\thefigure}{\arabic{figure}a}
      \centering
      \begin{minipage}{0.49\linewidth}
          \begin{figure}
              \epsfig{figure=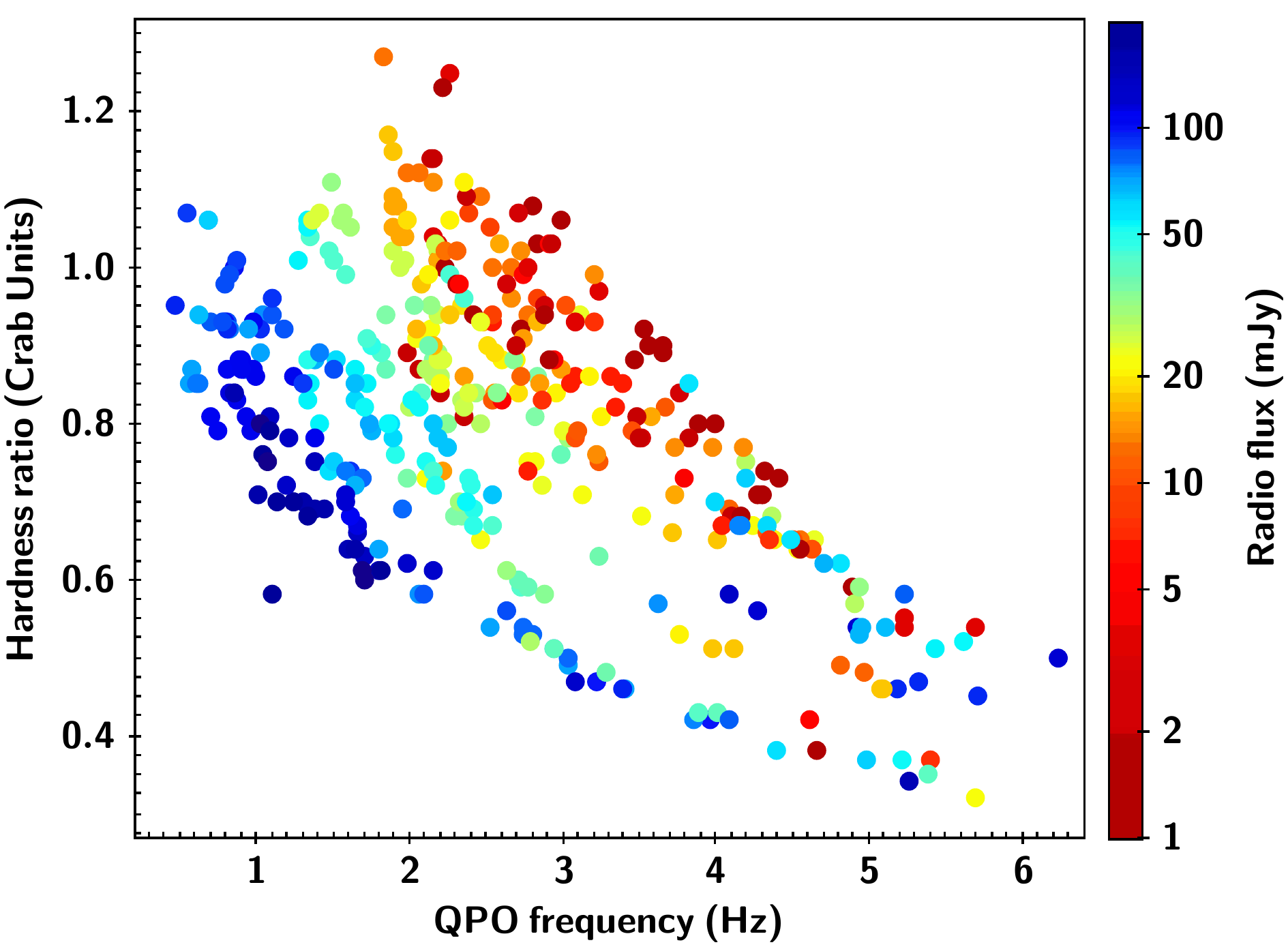,angle=0, height=5.5cm}
              \caption{}
              \label{fig:fig1a}
          \end{figure}
      \end{minipage}
\hfil
\setcounter{figure}{0}
\renewcommand{\thefigure}{\arabic{figure}b}
      \begin{minipage}{0.49\linewidth}
          \begin{figure}
              \epsfig{figure=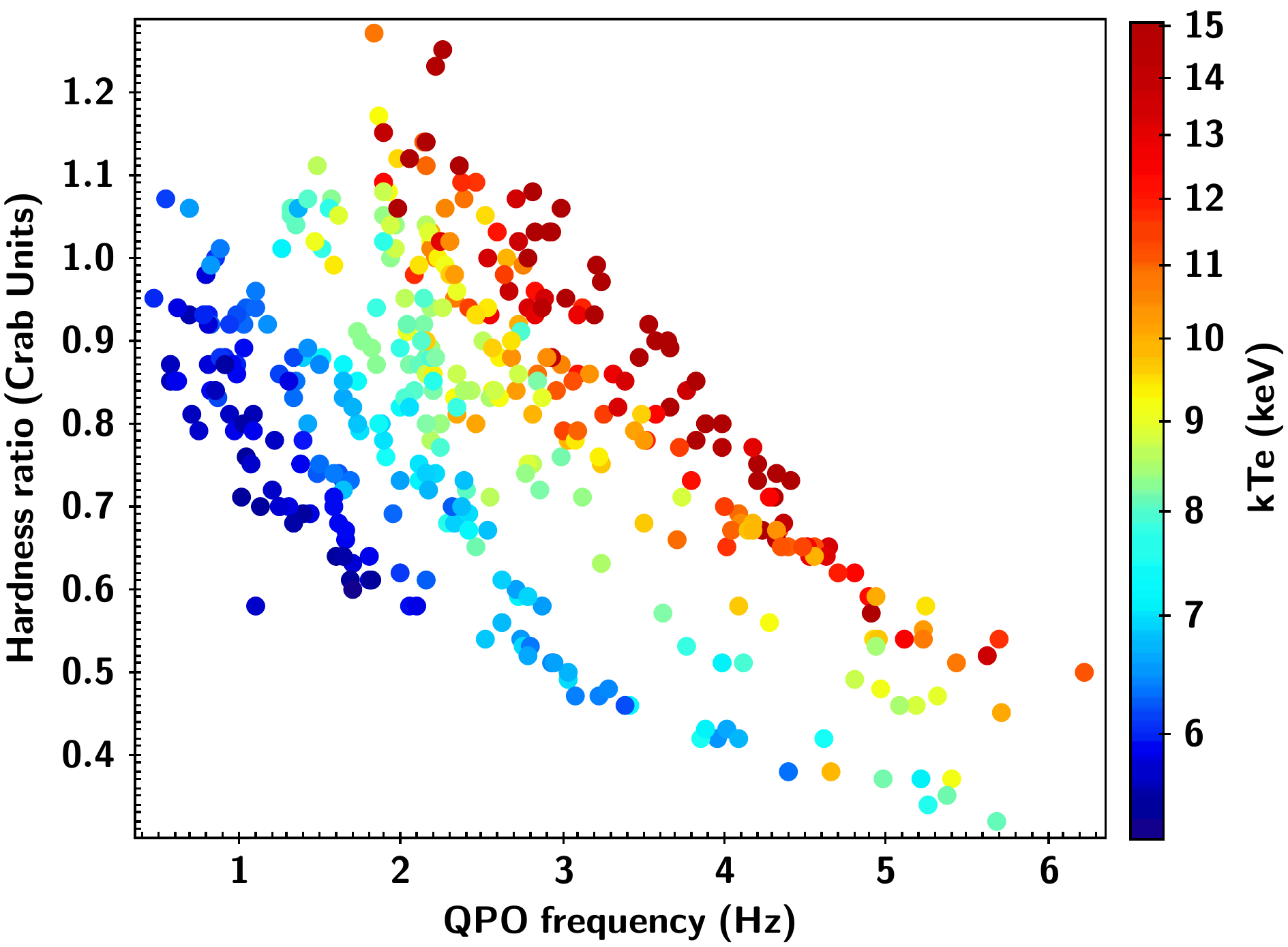, angle=0, height=5.5cm}
              \caption{}
              \label{fig:fig1b}
          \end{figure}
      \end{minipage}
  \end{minipage}

\setcounter{figure}{0}
\renewcommand{\thefigure}{\arabic{figure}c}
\begin{minipage}{\linewidth}
      \centering
      \begin{minipage}{0.49\linewidth}
          \begin{figure}
              \epsfig{figure=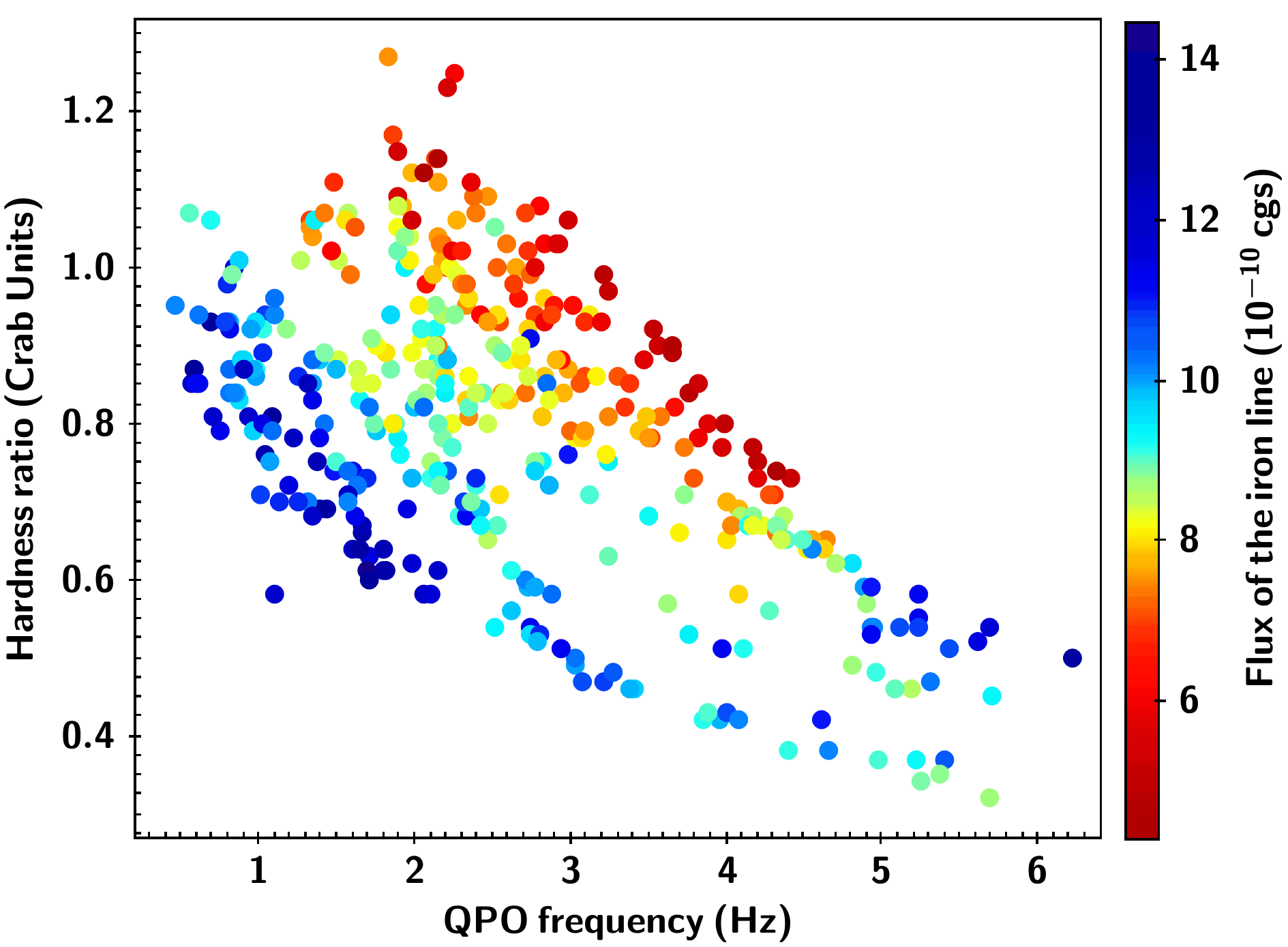, angle=0, height=5.5cm}
              \caption{}
              \label{fig:fig1c}
          \end{figure}
      \end{minipage}
\hfil
\setcounter{figure}{0}
\renewcommand{\thefigure}{\arabic{figure}d}
      \begin{minipage}{0.49\linewidth}
          \begin{figure}
              \epsfig{figure=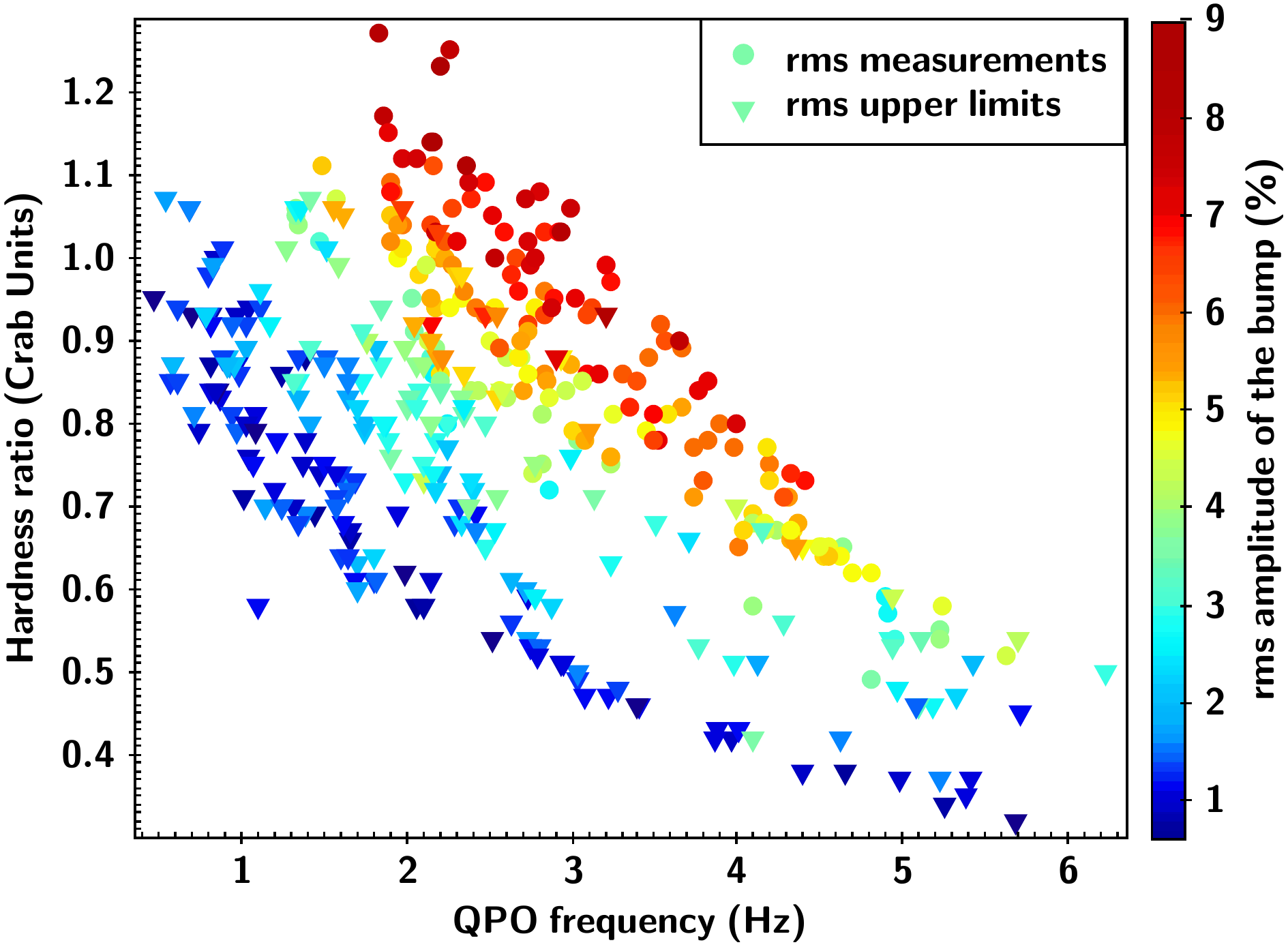, angle=0, height=5.5cm}
              \caption{}
              \label{fig:fig1d}
          \end{figure}
      \end{minipage}
  \end{minipage}
\renewcommand{\thefigure}{\arabic{figure}}

\renewcommand{\figurename}{}

\newpage
\setcounter{figure}{0}
\begin{figure}
\input{Fig1_caption.tex}
\end{figure}

\newpage
%!TEX root = GRS1915.tex 
\begin{figure}
\begin{center}
\epsfig{figure=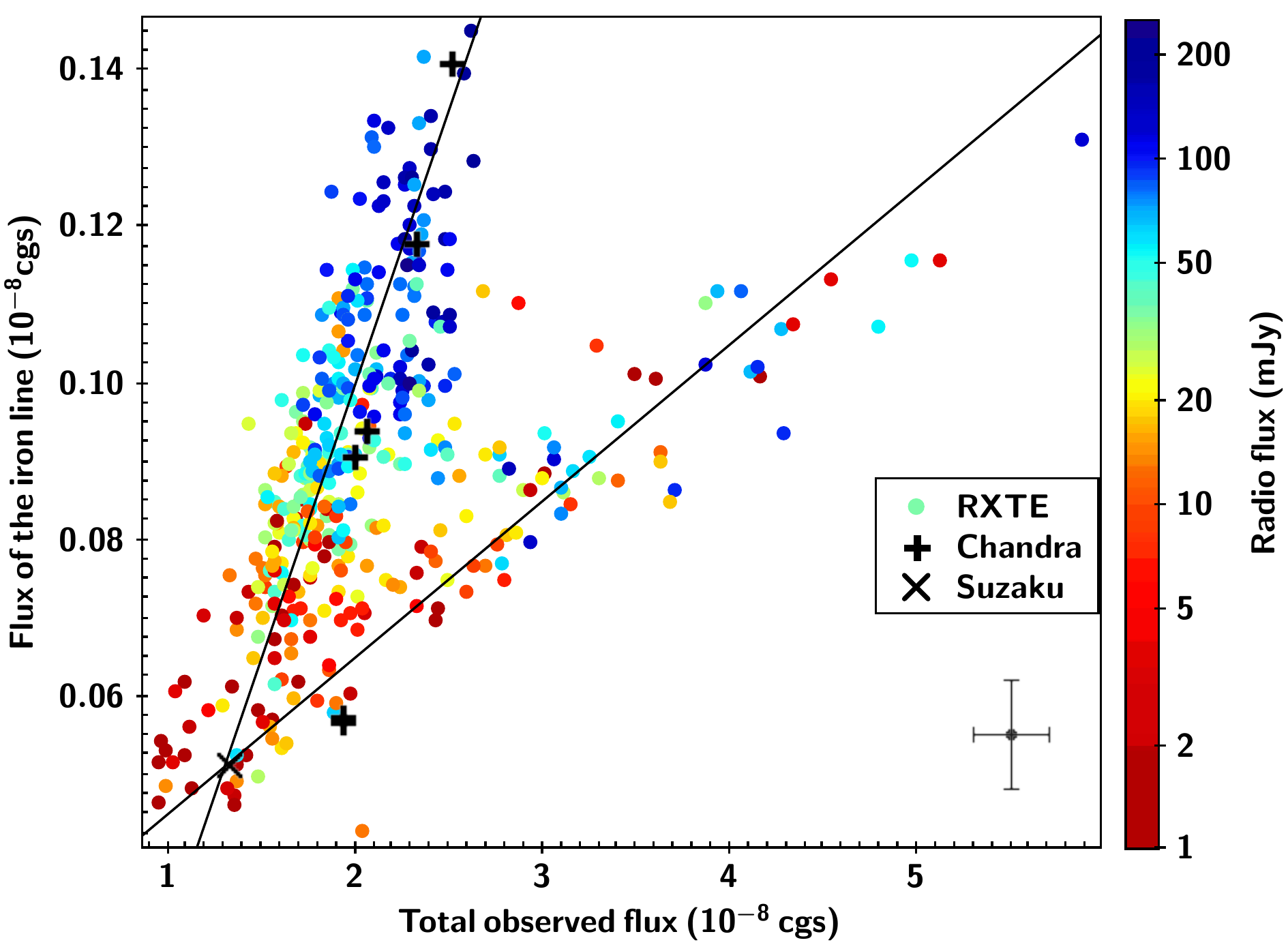, angle=0, height=5.5cm}
\hspace{0.5cm}
\epsfig{figure=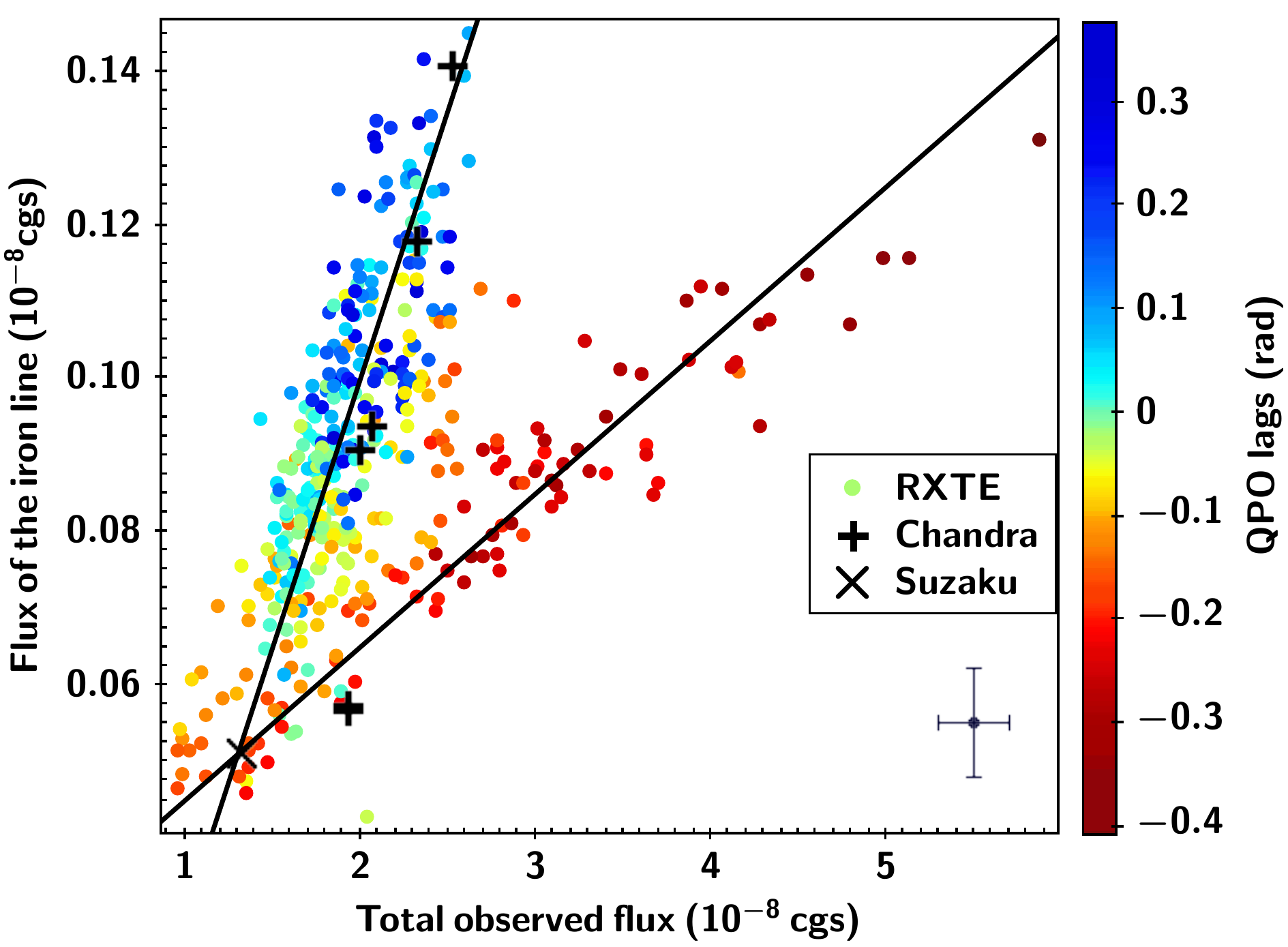, angle=0, height=5.5cm}
\end{center}
\end{figure}

\newpage
\setcounter{figure}{1}
\begin{figure}
\input{Fig2_caption.tex}
\end{figure}

\newpage
%!TEX root = GRS1915.tex 
\begin{figure}
\begin{center}
\epsfig{figure=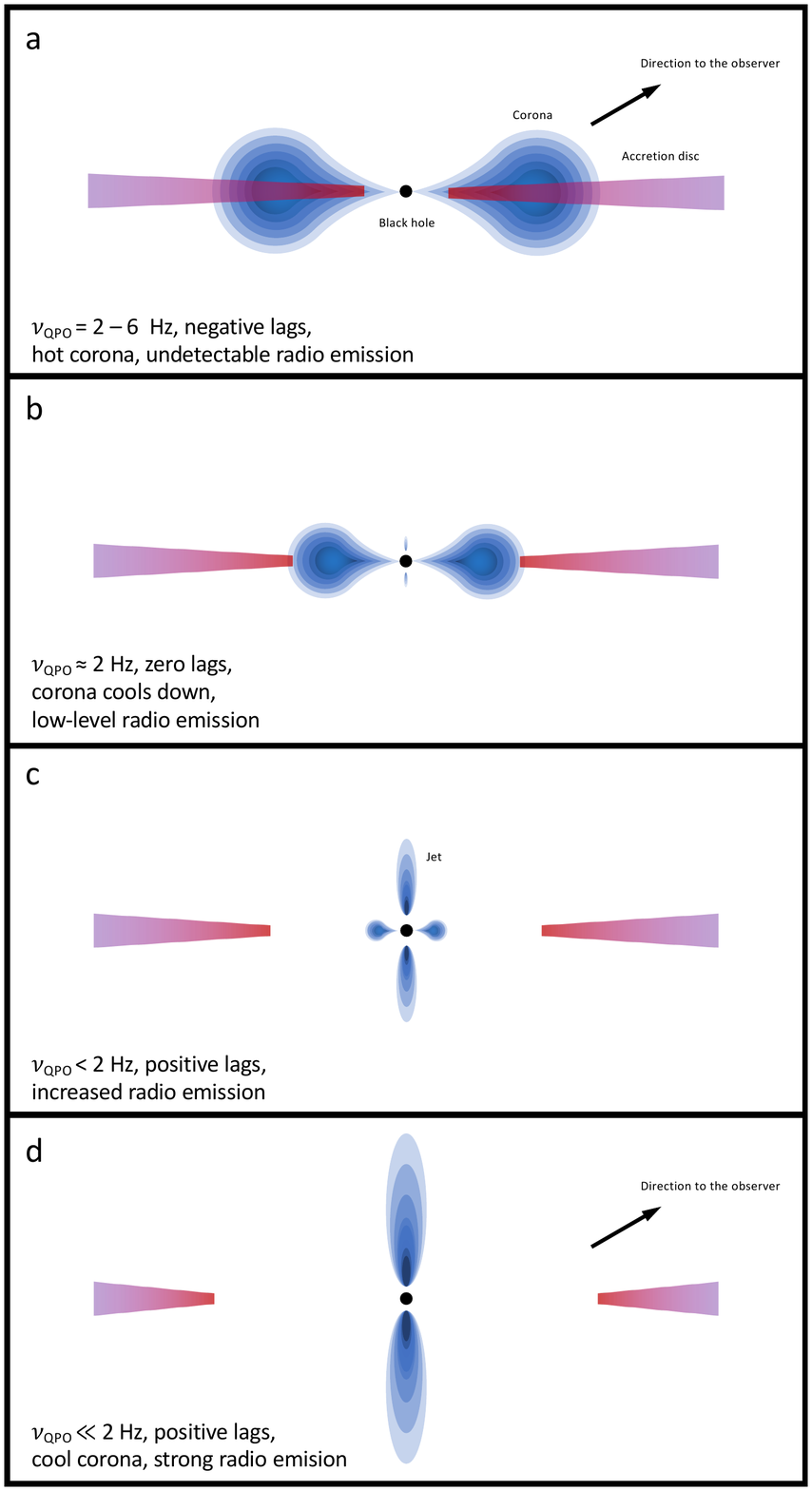, angle=0, width=15cm}
\end{center}
\end{figure}

\newpage
\setcounter{figure}{2}
\begin{figure}
\input{Fig3_caption.tex}
\end{figure}

\newpage
%!TEX root = GRS1915.tex 
\begin{figure}
\begin{center}
\epsfig{figure=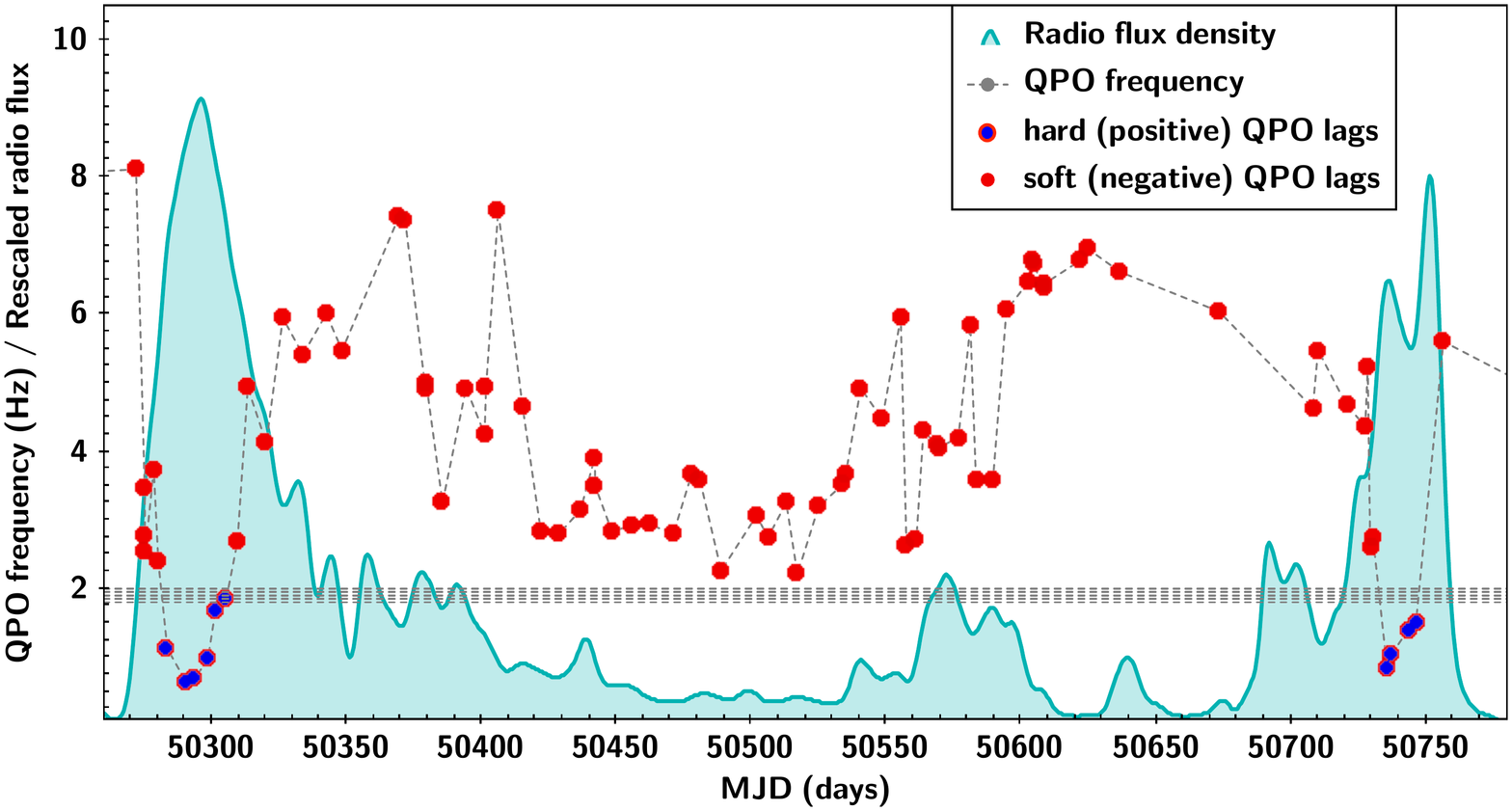,angle=90, width=15cm}
\end{center}
\end{figure}

\newpage
\setcounter{figure}{3}
\begin{figure}
\input{Fig4_caption.tex}
\end{figure}

\begin{linenumbers}

\newpage

%!TEX root = GRS1915.tex 
%
\begin{methods}
\label{sec:methods}

\subsection{Power spectra:}
\label{sec:methods-power}

We examined all the RXTE archival observations of GRS1915+105 from 1996 to 2012 obtained with the Proportional Counter Array (PCA). The observations that we used for our analysis belong to the class\autocite{Belloni-2000} $\chi$, state C, equivalent to one of the hard states in other (transient) black-hole sources.
For each observation we computed the Fourier power spectrum in the full energy band (absolute PCA channel $0$$-$$249$) every $128$ s with a time resolution of $1/512$ s, corresponding to a Nyquist frequency of $256$ Hz. We averaged all the $128$-s power spectra within an observation, subtracted the contribution due to Poisson noise\autocite{Zhang-1995} and normalised\autocite{Belloni-1990} these averaged power spectra to units of fractional rms squared per Hz. (We ignored the background count rate for this calculation since it was always negligible compared to the source count rate.) We subsequently applied a logarithmic frequency rebin to the data such that the size of a bin increases by $\exp{(1/100)}$ with respect to the size of the previous one, and we used XSPEC version 12.9 to fit the resulting power spectra with a sum of Lorentzian functions\autocite{Nowak-2000} that represent the broad-band noise component and a number of QPOs\autocite{Zhang-2020}. As in previous studies\autocite{Trudolyubov-2001} we included a Gaussian component centred at zero frequency in the model to fit a high-frequency bump at $60$$-$$80$ Hz in the power spectra. (We show two power spectra with the best-fitting model in Extended Data Figure~\ref{fig:power-spectra}.) For the rest of the analysis we selected only observations where at least one narrow QPO peak is present on top of the broad-band noise component in the power spectra, which is typical for the type-C QPOs\autocite{Casella-2005}. Based on the fitting results we only retained features that were detected at a significance greater than $3$-$\sigma$ and had a $Q$ factor, defined as the ratio of the QPO frequency to its width, of $2$ or more. We further checked the spectrogram of each observation, which shows visually the Fourier power spectrum as it varies with time, and excluded $\sim$$130$ observations in which the QPO frequency changed significantly within a full RXTE observation. Our final sample contains a total of $620$ observations. 

\subsection{Lag spectra:}
\label{sec:methods-lags}

Following the method described in\autocite{Vaughan-1997, Nowak-1999} we produced a frequency-dependent phase-lag spectrum (lag-frequency spectrum) between the $2$$-$$5.7$ and $5.7$$-$$15$ keV energy bands for each observation. Since our sample includes observations during the PCA calibration epochs $3$$-$$5$, to account for changes in the PCA channel-to-energy gain factor we selected the closest absolute channels that matched these energy bands, but the exact boundaries of each band still differ slightly between epochs. To calculate the phase lags of the QPO we averaged the lag-frequency spectra around the centroid frequency of the QPO, $\nu_{0} \pm FWHM/2$, where $FWHM~$ is the full-width at half-maximum of the Lorentzian that we used to fit the QPO profile. In principle, the phase lags in the range of frequencies of the QPO can be affected by the lags of the underlying broad-band noise component. However, in GRS 1915+105 the rms amplitude of the QPO is much higher than that of the broad-band noise and the phase lags at the frequency of the QPO are dominated by the QPO itself, with the contribution of the noise component being negligible\autocite{vandenEijnden-2016}. In this work, a positive (negative) lag means that the hard photons lag (lead) the soft photons. No correction for the dead-time-driven cross-talk\autocite{vanderKlis-1987} was done because this effect was found to be negligible. (See\autocite{Zhang-2020} for other details of the timing analysis.) In Extended Data Figure~\ref{fig:lag-spectra} we show nine representative power and lag-frequency spectra covering the range of frequencies of the QPO in GRS~1915+105. 

\subsection{Energy Spectra:}

We used the RXTE/PCA\footnote{We also fitted the RXTE/HEXTE (High Energy X-ray Timing Experiment) data of those observations in which the instrument was operational; the results were consistent with the ones of the RXTE/PCA, but since data of this instrument were not available for all observations, we did not use the RXTE/HEXTE data for the rest of the analysis.}
 Standard 2 data to extract energy spectra separately for each observation in our sample. We corrected the energy spectra for dead time and used the {\sc ftools} {\sc pcabackest} and {\sc pcarsp} in {\sc HEADAS} v.6.27 to, respectively, extract background spectra and produce response files for each observation. We fitted the energy spectra of all observations jointly with the model {\sc vphabs*(diskbb+gauss+nthcomp)}. The component {\sc vphabs} accounts for the interstellar absorption along the line of sight to the source; we used the abundance and cross-section tables given by\autocite{Wilms-2000} and \autocite{Verner-1996}, respectively, with the column density of hydrogen, $N_{\rm H}$, linked across observations and free to vary. Because $N_{\rm H}$ in the direction of GRS 1915+105 is quite high\autocite{Miller-2013}, $N_{\rm H}$$\simmore$$6$$\times$$10^{22}$ atoms cm$^{-2}$, we also let the abundance of iron in the interstellar material free to vary, since neutral iron produces an absorption edge at $E$$\approx$$7.1$ keV that was apparent in the fit residuals. The component {\sc diskbb}\autocite{Mitsuda-1984} represents the emission from an optically thick and geometrically thin accretion disc, and has parameters $kT_{\rm bb}$ and $N_{\rm dbb}$ that are, respectively, the temperature at the inner disc radius and the normalisation of the component defined as the ratio of the inner disc radius to the distance to the source squared multiplied by the cosine of the inclination of the disc with respect to the line of sight. The component {\sc nthcomp} \autocite{Zdziarski-1996} represents the emission due to inverse-Compton scattering from the corona. The parameters of this component are the power-law index, $\Gamma$, the electron temperature of the corona, $kT_{\rm e}$, the temperature of the source of soft photons that enter into, and are up scattered in, the corona, $kT_{\rm seed}$, and a normalisation that gives the flux density at $1$ keV. We assumed that the source of seed soft photons is the accretion disc, and linked the temperature $kT_{\rm seed}$ to the temperature of the {\sc diskbb} component for each observation separately during the fits. In this model, the optical depth of the corona, $\tau$, assumed to be homogeneous, is a function\autocite{Sunyaev-1980} of $\Gamma$ and $kT_{\rm e}$. Finally, the component {\sc gauss} represents a broad iron emission line at $6.5$$-$$7$ keV due to reflection of corona photons off the accretion disc\autocite{Fabian-1989, Fabian-2009}. While a full reflection model\autocite{Garcia-2014, Dauser-2014} would be more appropriate to describe this effect, the PCA instrument does not have enough spectral resolution to distinguish between the profile of a line in a full reflection model and a Gaussian component. The parameters of the {\sc gauss} component are the energy, width and total photon flux of the line, $E_{\rm g}$, $\sigma$, and $N_{\rm g}$, respectively. Finally, we calculated the total observed flux of the source and the intrinsic (i.e., not affected by interstellar absorption) flux of the full model and of all the individual model components in the $2$$-$$25$ keV energy range. In Extended Data Figure~\ref{fig:energy-spectra} we show the energy spectra, with the best-fitting model, of the same two observations in Extended Data Figure~\ref{fig:power-spectra}.

If the iron emission line is due to reflection of corona photons off the accretion disc, and the width of the line is a consequence of Doppler boosting, Doppler beaming and gravitational redshift, it would be more appropriate to use a full reflection component that includes relativistic broadening instead of a Gaussian to fit the spectrum.  To explore this we fitted all the spectra with the model {\sc vphabs*(diskbb+relxillCp+nthcomp)}, where the component {\sc rexillCp}\autocite{Dauser-2014} represents the reflection off the accretion disc of the emission from the {\sc nthcomp} component. Compared to the fits with a Gaussian, this model has six extra free parameters. Given that the RXTE/PCA instrument has limited spectral resolution and does not extend below $\sim$$3$ keV, the model is insensitive to some of the parameters of {\sc rexillCp}, and some of these parameters become highly correlated with those of the {\sc disc} component. We therefore fixed the spin of the black hole to $a_*=0.95$ (ref. \cite{Miller-2013}) and the inclination angle of the accretion disc to the line of sight to $i=65^{\circ}$ (ref. \cite{Mirabel-1994}). We further linked the two emissivity indices during the fits, which eliminates the parameter that gives the radius in the disc at which the power-law index of the emissivity profile changes, and fixed the inner radius of the disc at the radius of the innermost stable circular orbit\autocite{Miller-2013}. Finally we linked the power-law index, $\Gamma$, and the electron temperature, $kT_e$, of {\sc relxillCp} to the corresponding parameters of {\sc nthcomp}. The extra free parameters of the fit are one emissivity index, the reflection fraction, the ionisation parameter and the iron abundance of the disc. We obtain equally good fits as with the model above, with the reflection fraction ranging from 1\% to 20\%. Because the reflected component is much-lower than the Comptonised component, the parameters of the {\sc nthcomp} in these fits are consistent with those from the fits with a Gaussian line. The biggest impact of fitting the spectra with a reflection component is that the parameters of the disc are less well constrained in the model with {\sc rexillCp} than in the model with a Gaussian. This is understandable given that the emission of the reflection component at energies below $\sim$$4-5$ keV, where the disc emission peaks, is higher than that of a Gaussian. 

The {\sc rexillCp} model does not provide the flux or the equivalent width of the Gaussian, however the reflection fraction of {\sc relxillCp} is generally correlated with the parameters of the line\autocite{Dunn-2008}. We plotted the flux of the iron line from the fits with a Gaussian against the reflection fraction from the fits with {\sc relxillCp} and confirmed that this is indeed the case in our fits. We therefore obtain similar plots to those in Figure~\ref{fig:gauss-flux} if instead of the flux of the Gaussian line we plot the reflection fraction of {\sc relxillCp} vs. total flux. We prefer to use the former because, as we explained above, the model with a Gaussian line has less free parameters than the full reflection model which, given the limited spectral resolution of the RXTE/PCA instrument, leads to degeneracies of the parameters in the fits with the full reflection model, and because by using a Gaussian we can easily include in the plot the measurements of the flux of the iron line obtained with Chandra and Suzaku, as we did in Figure~\ref{fig:gauss-flux}.

\subsection{Hardness ratio:}
\label{sec:hardness}

For each RXTE observation of GRS 1915+105 we computed a hardness ratio value defined as the ratio of the background-subtracted count rate of the source in the $13$$-$$60$ keV band to that in the $2$$-$$7$ keV band. As with the lag-frequency spectra, we selected the closest absolute channels that matched these energy bands in each PCA gain epoch. Before calculating the ratios, we corrected each observed count rate for instrumental dead time and normalised them by the count rate in the same band from the Crab nebula to account for possible changes of the effective area of the instrument with time. 

\subsection{Radio fluxes:}
\label{sec:radio}

For the flux-density data of GRS 1915+105 in radio we used measurements\autocite{Pooley-1997} from the Ryle Telescope at $15$ GHz with the four mobile antennas set in a compact array configuration within $100$ m of the nearest fixed antenna yielding a resolution of $\sim$$30$ arcsec at that frequency. The flux-density scale was calibrated with observations of the nearby quasars 3C 48 or 3C 286. The observations consist mostly of  32-s samples with an rms noise of $6$ mJy that decreases as the square root of the integration time; flux-density values below about $1$ mJy may be unreliable. See\autocite{Pooley-1997} for other details of the analysis of the radio data. 

Finally, we cross-correlated all the X-ray and radio data based on the date of the observations, which left us with a sample of $410$ observations with simultaneous radio flux densities at 15 GHz and X-ray energy, power-density and lag-frequency spectra and hardness ratios. 

Having described the observations and analysis we used, it is worth noting that the measurements presented in this paper come from very different types of data and totally independent analysis techniques: The hardness ratio, iron-line flux and $kT_{\rm e}$ come from X-ray light curves and time-averaged energy spectra, the frequency and lags of the QPO and the rms amplitude of the high-frequency bump come from Fourier power spectra of high-time resolution data, and the radio flux was measured independently in a totally different frequency band than the X-ray data.

\end{methods}

\newpage
\begin{extended-discussion}

\subsection{Fits to the power spectra:}

In Extended Data Figure~\ref{fig:power-spectra} we plot two representative Fourier power spectra of GRS~1915+105 to show both the type-C QPO and the high-frequency bump. The power spectrum plotted in black corresponds to the observation from September 21 1998 (ObsID 30703-01-34-00), while the power spectrum plotted in blue corresponds to the observation from April 10 1998 (ObsID 30402-01-09-01). Both power spectra were calculated using photons from the full PCA band, nominally $2$$-$$60$ keV. The QPO is the narrow feature that appears in both power spectra at $\sim$$2$ Hz, with a second harmonic and subharmonic at, respectively, $\sim$$4$ Hz and $\sim$$1$ Hz. 
In the case of GRS~1915+105 it is relatively straightforward to distinguish the QPO fundamental from the subharmonic and harmonic components because the QPO fundamental, subharmonic and harmonic components show very different rms amplitude and lags dependence when plotted vs. their own frequency\autocite{Zhang-2020}, and the fundamental is always the strongest peak in the power spectrum.

The model includes also two broad Lorentzians centred, respectively, at  $0$ Hz and $2.5$ Hz to fit the the broad-band noise component in the power spectrum. The high-frequency bump is the broad feature extending up to $\sim$$100$ Hz in the black power spectrum\autocite{Belloni-1999, Revnivtsev-2000, Nowak-2000, Belloni-2002, Pottschmidt-2003, Motta-2014a, Motta-2014b}. We fitted this bump with a zero-centred Gaussian and let the width and normalisation free to vary during the fits. It is apparent that the QPO at 2 Hz and its second harmonic and subharmonic are present in both power spectra, but the high-frequency bump is only present in the observation of September 1998 and significantly absent in the one of April 1998. We computed the rms amplitude of the high-frequency bump by integrating the power of the best-fitting Gaussian from $0$ Hz to $\infty$. The rms amplitude of the high-frequency bump in the observation of September 1998 is $8.0$$\pm$$0.3$\% of the total luminosity of the source in the $2$$-$$60$ keV band, whereas the 95\% confidence upper limit of the rms amplitude in the observation of April 1998 is only $1$\% of the total luminosity. Since the high-frequency bump was not significantly detected in the observation of April 1998, to calculate the upper limit we fixed the width of the Gaussian to the value we obtained from the fit to the observation of September 1998. We followed the same procedure to compute the rms amplitude or, when appropriate, the upper limits of the high-frequency bump in the rest of the observations. In the observations in which the high-frequency bump is significantly detected the width of this feature ranges from $\sim$$30$ Hz to $\sim$$100$ Hz. In those observations in which the high-frequency bump was not significantly detected, to compute the upper limits we fixed the width of the Gaussian to $70$ Hz, which is the average value that we obtained from the fits of the observations in which the high-frequency bump was significantly detected. We report a detailed analysis of the properties of the high-frequency bump in a separate paper\autocite{Zhang-2021}.

\subsection{Fits to the energy spectra:}
\label{fit-energy}

In Extended Data Figure~\ref{fig:energy-spectra} we show the energy spectra of the same two observations for which we show the power spectra in Extended Data Figure~\ref{fig:power-spectra}. The spectrum plotted in black corresponds to the observation from September 21 1998 (ObsID 30703-01-34-00), while the spectrum plotted in blue corresponds to the observation from April 10 1998 (ObsID 30402-01-09-01). This Figure shows that the flux of the source was lower and, at the same time, the spectrum was harder (extending to higher energies) in September 1998 than in April 1998. This difference is reflected in the values of the best-fitting temperature of the corona in these two observations, which are $kT_{\rm e}$$=$$10.9$$\pm$$0.8$ keV and $kT_{\rm e}$$=$$6.3$$\pm$$0.5$ in the observations of September 1998 and April 1998, respectively. From the comparison of this Figure and Extended Data Figure~\ref{fig:power-spectra} it is apparent that the high-frequency bump is present when the temperature of the corona is high. Indeed, when we fit all the observations we find that the rms amplitude of the high-frequency bump is higher when the temperature of the corona is higher, regardless of the values of the hardness ratio and QPO frequency in each observation (see Figure~\ref{fig:fig1}).

\subsection{Energy-dependent rms amplitude and phase lags of the QPO. The corona becomes the jet:}

In Extended Data Figure~\ref{fig:lag-spectra} we plot the power- and lag-frequency spectra of nine observations of GRS~1915+105, representative of the range of frequencies covered by the type-C QPO. The phase lags of the QPO change from positive when the frequency of the QPO is lower than $\sim$$2$ Hz to negative when the frequency of the QPO is higher than $\sim$$2$ Hz. (See\autocite{Zhang-2020} for details of this result and a discussion of the implication of this behaviour.) At the same time, as the QPO frequency increases the slope of the lag spectrum decreases\autocite{Karpouzas-new}, going from positive when the frequency increases from from $\sim$$0.5$ Hz to $\sim$$2$ Hz, to $0$ when the frequency is $\sim$$2$ Hz, and finally to negative when the QPO frequency increases further from $\sim$$2$ Hz to $\sim$$6$ Hz. 

The energy-dependent rms amplitude and phase lags of the QPO in GRS 1915+105 are well explained by a time-dependent model\autocite{Karpouzas-2020} in which photons from the accretion disc are inverse Compton scattered in a corona of characteristic size $L$, and a fraction of those photons subsequently impinges back onto the disc before reaching the observer. The time-averaged version of this variability model is equivalent to the {\sc nthcomp} model in XSPEC.
In the current version of the model we assume that the corona is spherically symmetric and has constant optical depth and electron temperature. Most likely the geometry of the corona is more complex than that, and the optical depth and electron temperature change as a function of the distance to the black hole. In a non-spherical corona the rms amplitude and lags of the QPO may change with the inclination of the binary system with respect to the line of sight\autocite{Motta-2015,vandenEijnden-2017}. We note, however, that both the rms amplitude and lags of the QPO in this and other sources change in a systematic way with QPO frequency, which in turn changes over time scales of minutes to hours, much faster than potential changes of the binary inclination that are expected to happen over periods of a few tens of days to a few thousand years\autocite{Wijers-1999}. This evinces that, although inclination may play a role, the thermodynamical properties of the corona have a much larger effect than inclination in setting the rms amplitude and lags of the QPO.

Fits with this time-dependent model to the data of GRS~1915+105 show\autocite{Karpouzas-new, Garcia-new} that, as the QPO frequency decreases from $\sim$$6$ Hz to $\sim$$2$ Hz, the feedback fraction is close to 100\% and remains more or less constant. At the same time the size\footnote{In the model we solve the Kompaneets equation\autocite{Kompaneets-1957} for a spherical corona; since the actual geometry of the corona is likely different, the values given here should be considered as a characteristic size of the corona.} of the corona decreases from $L$$\approx$$1000$$-$$2000$ km ($\sim$$60$$-$$120$ $R_g$ for a 12 solar-mass black hole) to $L$$\approx$$100$ km ($\sim$$6$$-$$10$ $R_g$), the lags are negative and their magnitude decreases. 
At a QPO frequency of $\sim$$2$ Hz the corona size is minimum, the feedback fraction drops abruptly to zero and the lags become zero. 
If the QPO frequency is produced by Lense-Thirring precession\autocite{Stella-1998, Ingram-2009} of the inner edge of the accretion disc, at a QPO frequency of 2 Hz the size of the corona is more or less equal to the inner disc radius (dashed line in Extended Data Figure~\ref{fig:size}). When the QPO frequency decreases below $\sim$$2$ Hz, the corona size increases again up to $L$$\approx$$2$$\times$$10^{4}$ km ($\sim$$1200$ $R_g$) at the lowest QPO frequency, 
the feedback fraction remains constant close to zero and the lags become positive and increase with decreasing QPO frequency. Although the size of the corona increases again when the QPO frequency is below $\sim$$2$ Hz, the fact that the feedback fraction is close to zero whereas when the QPO frequency is above $\sim$$2$ Hz the feedback fraction is high, implies that the corona does not cover the inner parts of the accretion disc. This in turn demonstrates that the geometry of the corona has changed (see Fig.~\ref{fig:fig3c}).

Extended Data Figure~\ref{fig:radio-qpo} shows the time evolution of the QPO frequency superimposed on 10 years of the radio flux measurements of GRS~1915+105 with the Ryle telescope. To produce this Figure we reanalysed all the RXTE observations in the period of the radio monitoring campaign, and added 357 observations with QPOs that were not in the sample of 410 observations discussed so far. These 357 observations were not included in the original analysis\autocite{Zhang-2020} because the QPO frequency drifts by a few tenths of Hz in the observation period, which could bias the measurement of the rms and lags of the QPO. But since the frequency drift is only slightly larger than the error of the average QPO frequency in each observation, we decided to use these data for this Figure because this allows us to have a denser coverage of the time evolution of the QPO frequency during the radio monitoring campaign. Extended Data Figure~\ref{fig:radio-qpo} shows that the intervals of strong radio flares\footnote{While it is apparent to the eye from the Figure, it is difficult to quantify what a strong radio flare is. Here we classify a flare as strong if its flux density exceeds 50 mJy for more than 10 days and the flare lasts more than 100 days.} coincide with the times when the QPO frequency is below 2 Hz.
This behaviour repeats consistently for all the radio flares in these observations; there is no radio flare in which the QPO frequency is not below 2 Hz, and there is no case of a QPO frequency below 2 Hz without a radio flare. Given the data, we estimated that the probability of having a QPO frequency below 2 Hz and a radio flare at the same time if the two phenomena were uncorrelated is less than $2$$\times$$10^{-10}$.

Taken together, the results shown in Extended Data Figures~\ref{fig:size} and \ref{fig:radio-qpo} and the sudden change of the feedback fraction\autocite{Karpouzas-new, Garcia-new} from $\sim$$1$ to $\sim$$0$ when the QPO frequency is, respectively, above or below 2 Hz are consistent with the scenario described earlier, in which the energy that is initially stored in the X-ray corona is gradually released into the radio jet and, quite possibly, the X-ray corona itself becomes the jet.

The morphing of the extended corona into the jet is strengthened by the relation between the actual luminosity of the corona and that expected if the emission is due to inverse-Compton scattering of seed photons from the accretion disc. In Extended Figure~\ref{fig:compton-y} we show the ratio of the expected to the observed corona flux vs. the observed flux of the corona. We calculated the expected flux as the product of the bolometric disc flux times $e^{y}$, where $y$ is the Compton parameter. (This is the expression for the case that the optical depth is $\tau > 1$.) Since the Compton $y$ parameter gives the number of scatterings times the relative energy that a photon gains per scattering, if the seed-photon source is the disc, the expected and observed corona fluxes should be roughly the same. This Figure shows that the observed and expected fluxes are the same (within a factor $3$) only when the lags of the QPO are negative (yellow, orange and red points), i.e., when the QPO frequency is higher than $\sim$$2$ Hz and the disc is enshrouded by the corona. When the lags are positive (light blue and blue points) and the corona morphed into the jet, if the seed photons were from the disc the expected flux of the corona is much lower than the one observed. In this phase a different source must be providing the seed photons for Comptonisation. Since in this phase the jet is active, the most likely source of seed photons is synchrotron emission\autocite{Veledina-2018} from the same electrons that produce the jet. Indeed, if we calculate the expected corona flux by multiplying the observed synchrotron flux coming from the jet extrapolated to the X-ray band times $e^{y}$, the expected corona flux is much higher when the lags of the QPO are positive than when they are negative.

\end{extended-discussion}

\subsection{Data availability}

All the X-ray data used in this study are available NASA's High Energy Astrophysics Science Archive Research Center  \url{https://heasarc.gsfc.nasa.gov/}. The radio data used in this study are available through the website \url{http://www.mrao.cam.ac.uk/~guy/}.

\subsection{Code availability}

The data reduction was done using HEADAS v6.27, while the model fitting of energy, power and lag-energy spectra was done with XSPEC; both packages are available at the HEASARC website (\url{https://heasarc.gsfc.nasa.gov/}). The timing analysis was performed with the GHATS package developed by T.M.B. and is available on request (\url{http://astrosat.iucaa.in/~astrosat/GHATS_Package/Home.html}). All figures were made in TOPCAT, a JAVA-based scientific plotting package developed by Mike Taylor (\url{http://www.star.bris.ac.uk/~mbt/topcat/}).

\end{linenumbers}

\newpage
\begin{extended}

\renewcommand{\figurename}{\bf Extended Data Figure}    

\setcounter{figure}{4}
%!TEX root = GRS1915.tex 
\begin{figure}
\centerline{\epsfig{figure=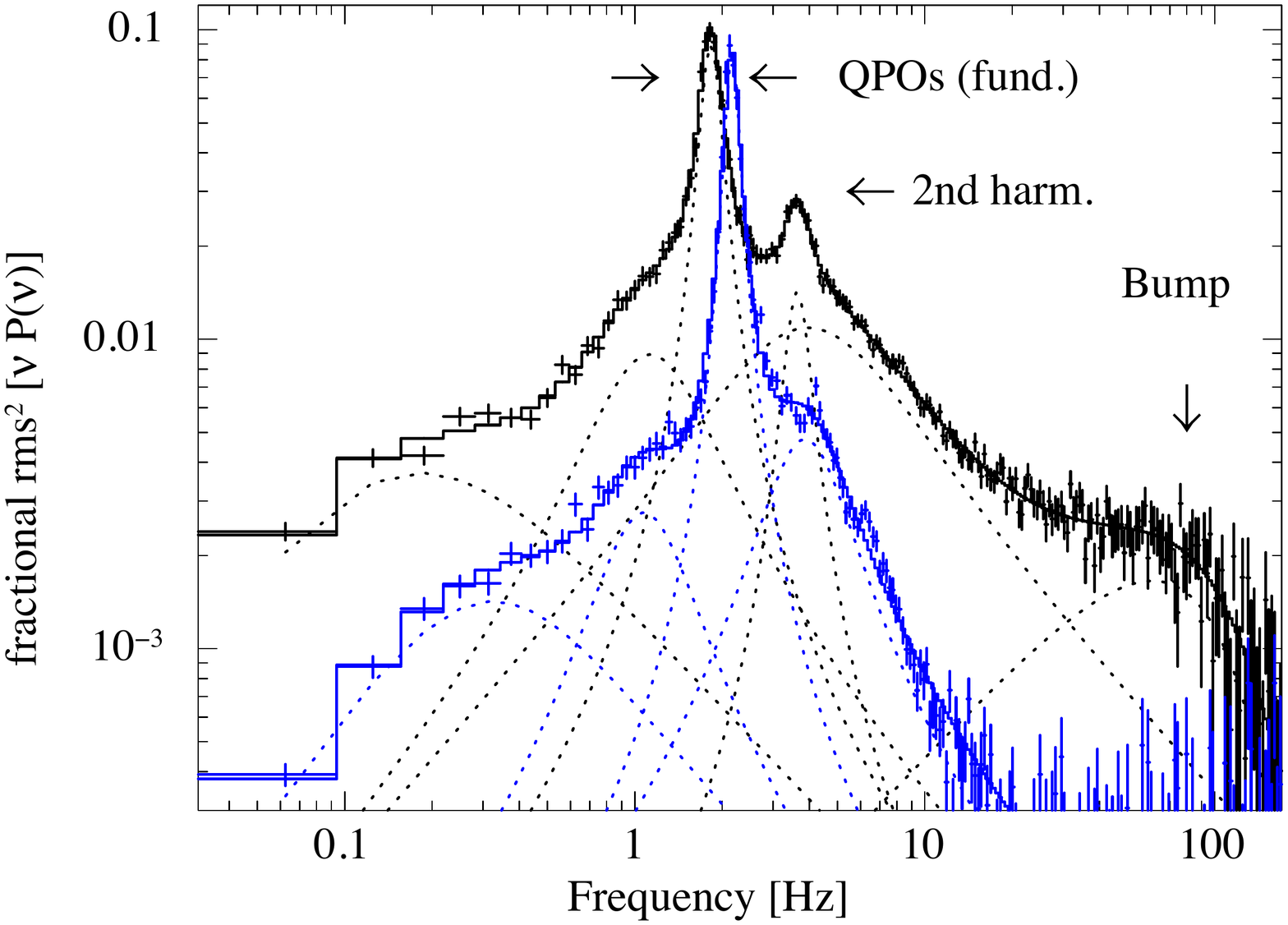,angle=-90, width=20cm}}
\vspace{-1cm}
\end{figure}

\newpage
\begin{figure}
\input{Fig5_caption.tex}
\end{figure}

\newpage
\setcounter{figure}{5}
%!TEX root = GRS1915.tex 
\begin{figure}
\centerline{\epsfig{figure=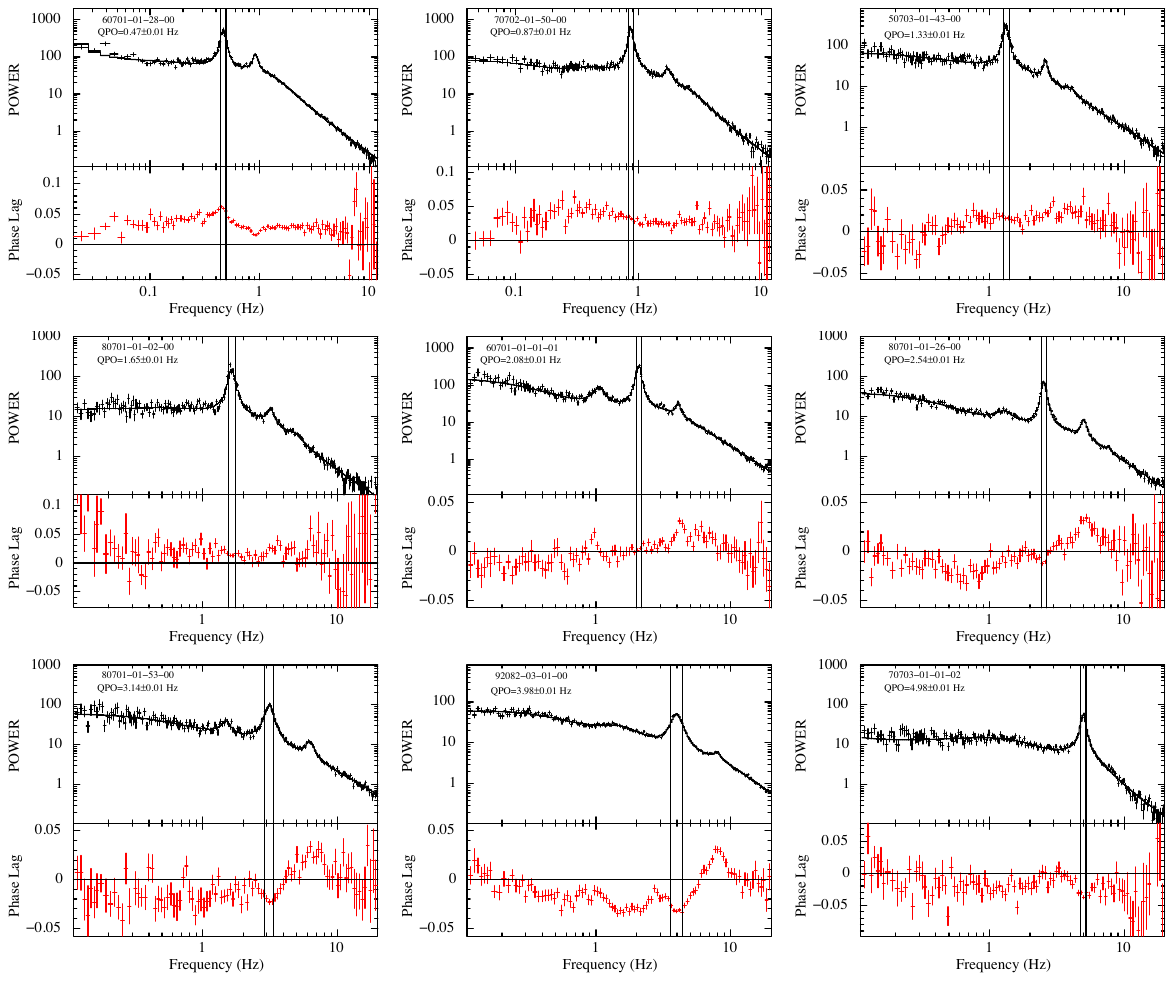,angle=0, trim=0cm 9cm 0cm 8cm, clip, width=27cm}}
\vspace{-1cm}
\end{figure}

\newpage
\begin{figure}
\input{Fig6_caption.tex}
\end{figure}

\newpage
%!TEX root = GRS1915.tex 
\begin{figure}
\centerline{\epsfig{figure=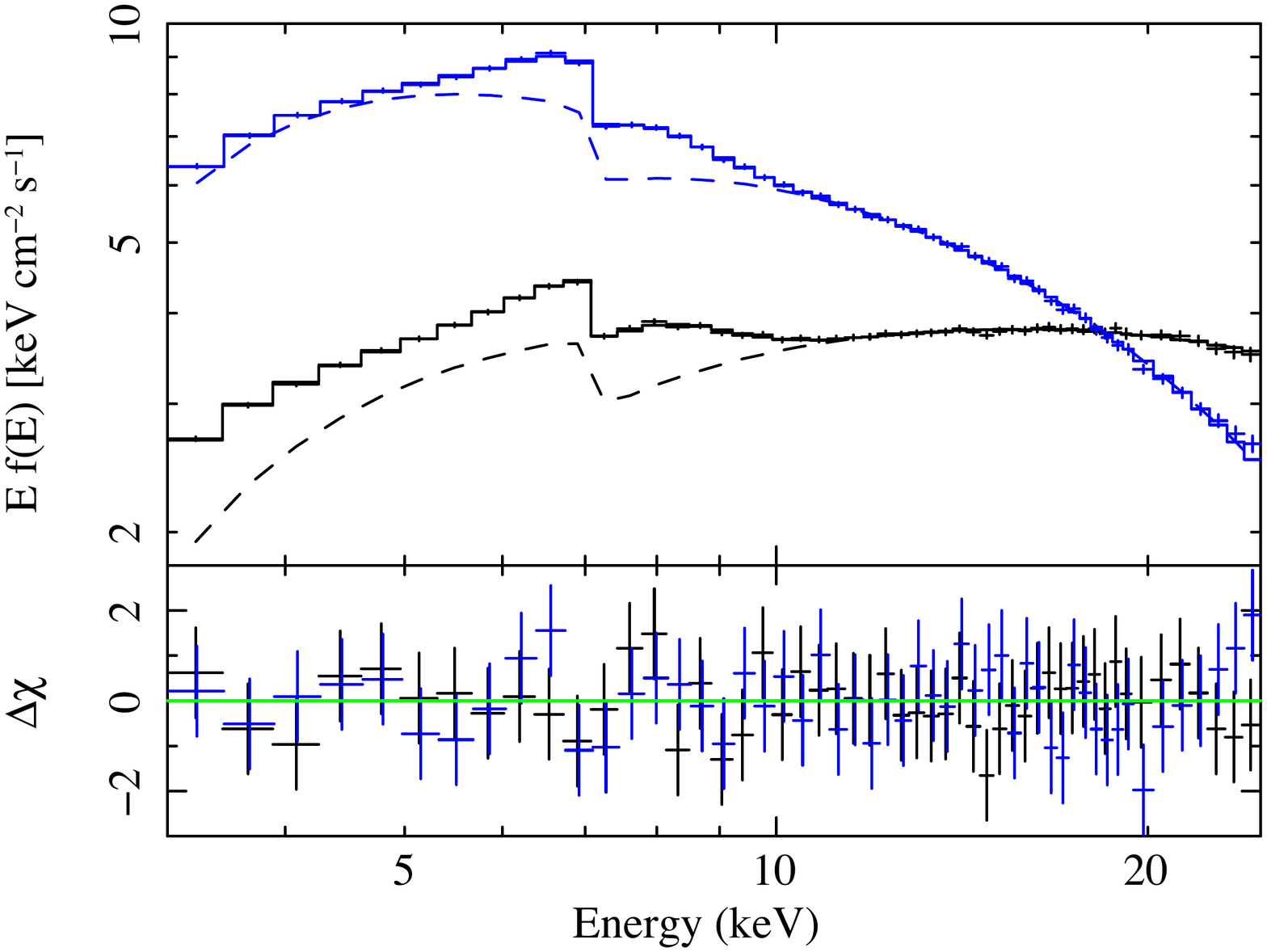,angle=-90, width=20cm}}
\vspace{-1cm}
\end{figure}

\newpage
\setcounter{figure}{6}
\begin{figure}
\input{Fig7_caption.tex}
\end{figure}

\newpage
%!TEX root = GRS1915.tex 
\begin{figure}
\begin{center}
\epsfig{figure=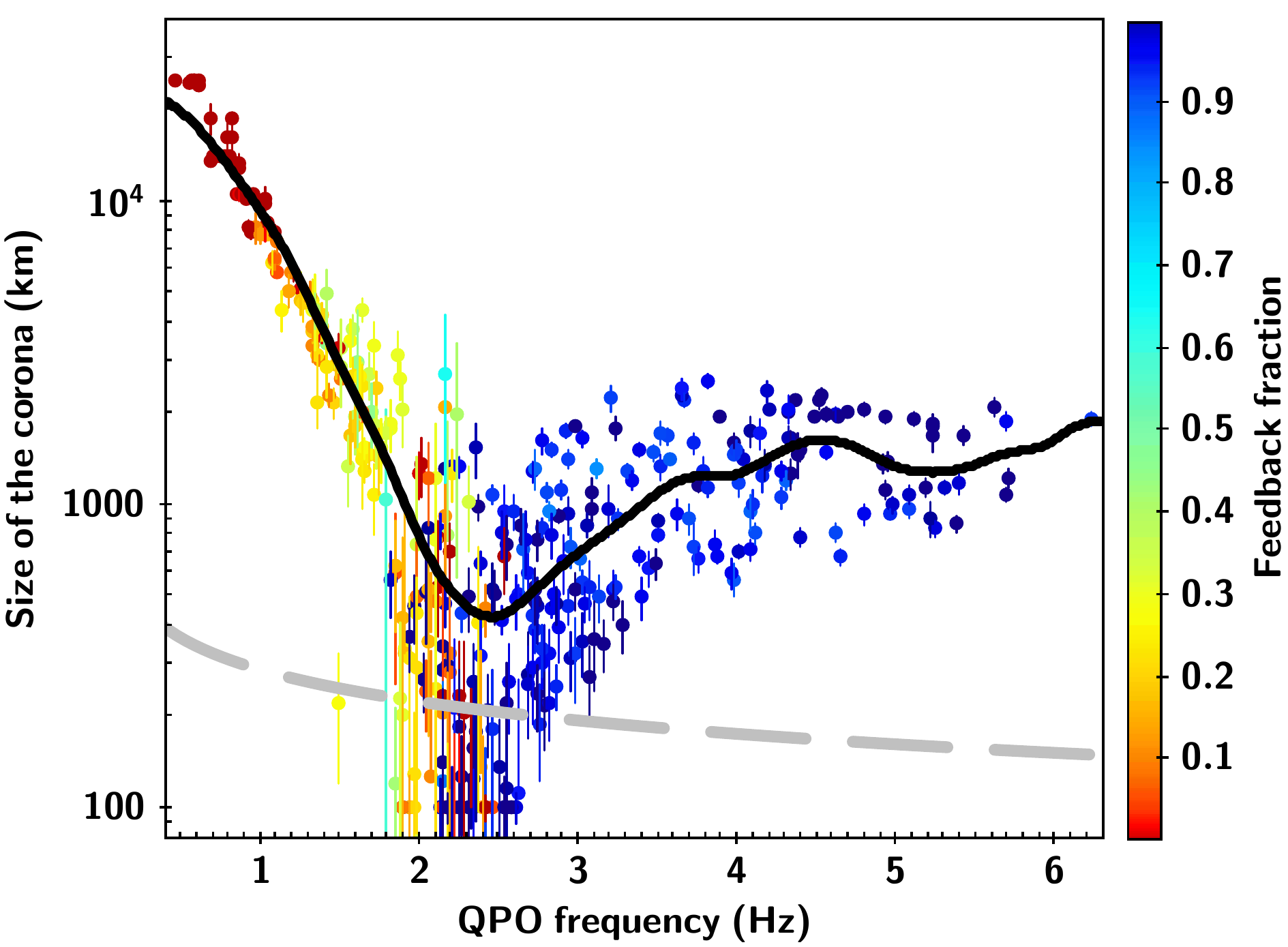, angle=0, width=18cm}
\end{center}
\end{figure}

\newpage
\setcounter{figure}{7}
\begin{figure}
\input{Fig8_caption.tex}
\end{figure}

\newpage
%!TEX root = GRS1915.tex 
\begin{figure}
\begin{center}
\epsfig{figure=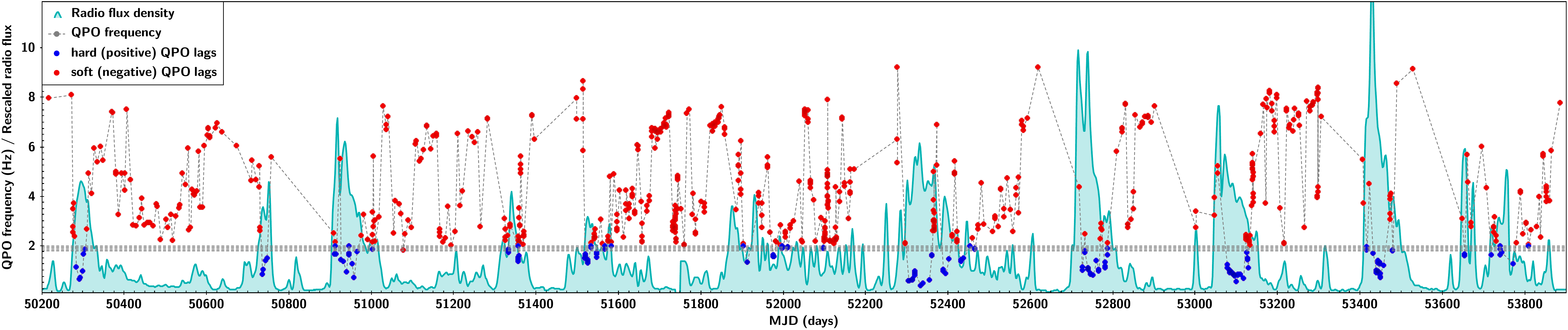, angle=-90, width=4.5cm}
\end{center}
\end{figure}

\newpage
\setcounter{figure}{8}
\begin{figure}
\input{Fig9_caption.tex}
\end{figure}

\newpage
%!TEX root = GRS1915.tex 
\begin{figure}
\begin{center}
\epsfig{figure=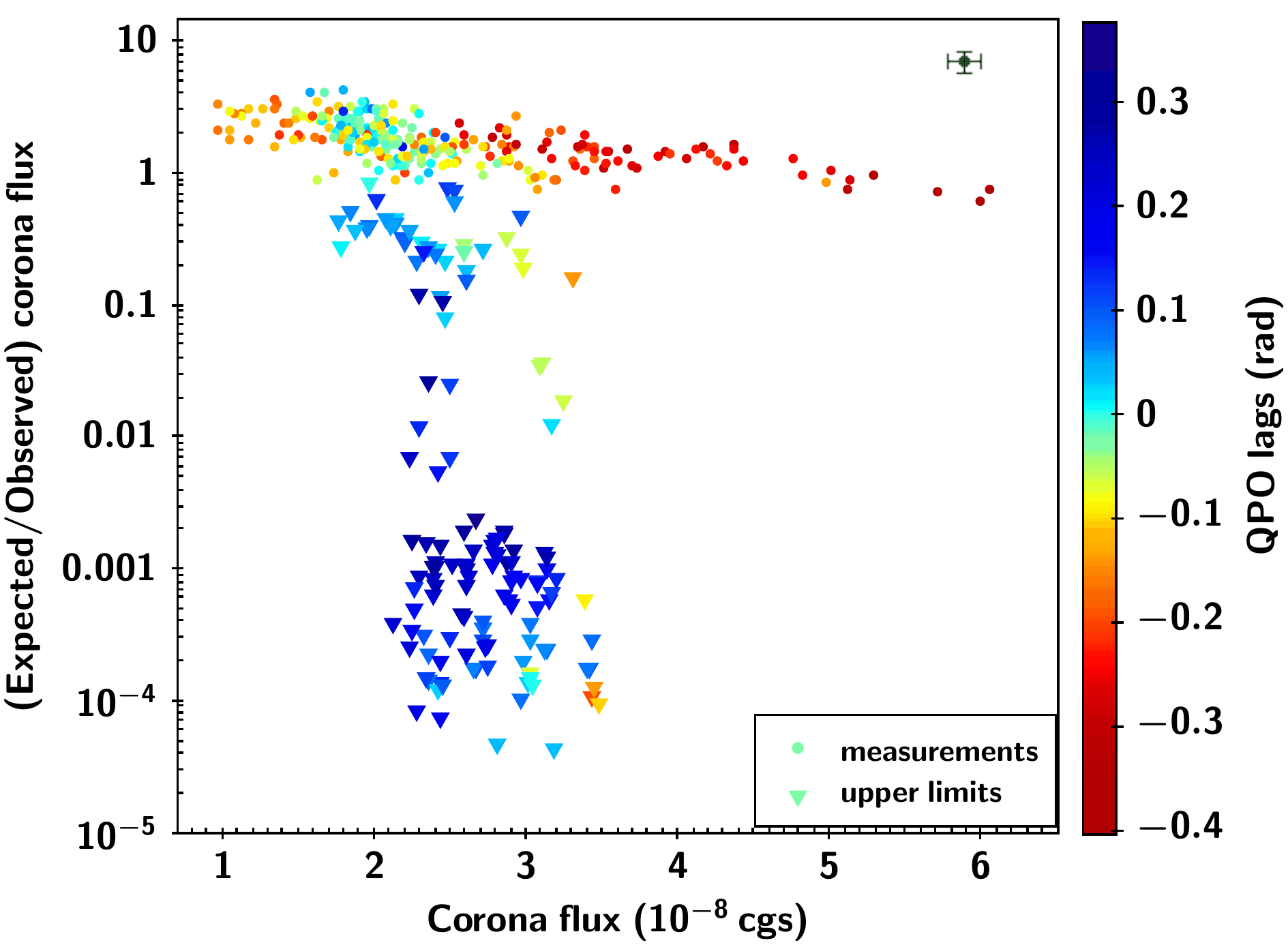,angle=0, width=15cm}
\end{center}
\end{figure}

\newpage
\setcounter{figure}{9}
\begin{figure}
\input{Fig10_caption.tex}
\end{figure}

\newpage

\end{extended}

\newpage
\printbibliography[segment=\therefsegment,check=onlynew]

\end{document}